\documentclass[letterpaper, 10 pt, conference]{ieeeconf}
\IEEEoverridecommandlockouts
\usepackage[utf8]{inputenc}
\usepackage{amsmath}
\usepackage{graphicx}
\usepackage{amssymb}
\usepackage{url}
\usepackage{bm}
\usepackage{comment}
\usepackage[dvipsnames]{xcolor}
\usepackage[english]{babel}
\usepackage{tikz}
\usepackage{bm}
\usepackage{mathtools}
\usepackage{matlab-prettifier}
\usepackage{cuted}
\usepackage{algorithm, algorithmic}
\usepackage{caption}

\graphicspath{{Figures}}

\newtheorem{lemma}{Lemma}

\title{\textbf{MPC for Tracking applied to rendezvous with non-cooperative tumbling targets ensuring stability and feasibility}}

\author{{Jose Antonio Rebollo, Rafael Vazquez,  Ignacio Alvarado and Daniel Limon}
\thanks{J.A. Rebollo and R. Vazquez are with the Department of Aerospace Engineering, Universidad de Sevilla, Camino de los Descubrimiento s.n., 41092 Sevilla, Spain.}%
\thanks{I. Alvarado and D. Limon are with the Department of Systems Engineering and Automation, Universidad de Sevilla, Camino de los Descubrimiento s.n., 41092 Sevilla, Spain.}%
}

\begin{document}

\maketitle
 \interdisplaylinepenalty=2500


\baselineskip=.9 \normalbaselineskip

\begin{abstract}
A Model Predictive Controller for Tracking is introduced for rendezvous with non-cooperative tumbling targets in active debris removal applications. The target's three-dimensional non-periodic rotational dynamics as well as other state and control constraints are considered. The approach is based on applying an intermediate coordinate transformation that eliminates the time-dependency due to rotations in the constraints. The control law is then found as the solution to a QP problem with linear constraints and dynamics, as derived from the HCW equations, that provides feasibility and stability guarantees by means of a terminal LQR and dead-beat region. The proposed control algorithm performs well in a realistic simulation scenario, namely a near rendezvous with the Envisat spacecraft.
\end{abstract}

\section{Introduction}\label{Sec1}
Space debris constitutes a significant and growing threat to operational spacecraft in Earth's vicinity. As the number of launches and satellite deployments has recently increased, the potential for in-orbit collisions has grown, prompting the need for Active Debris Removal (ADR) strategies. ADR aims to remove defunct or non-operational objects from Earth's orbit, mitigating the risk of collisions with operational spacecraft and reducing the overall debris population \cite{Sha16}.

Several rendezvous-based ADR technologies have been proposed \cite{MARK2019194}. These strategies require a mechanical attachment to the target in order to place a deorbiting device such as Electrodynamic Tethers (ET) \cite{San20d,Sar21a} or solid propellant kits, among others \cite{CASTRONUOVO2011848}. However, one of the primary challenges in ADR is the safe and precise rendezvous with non-cooperative targets, that is, objects that are not designed for docking and may even be tumbling in space \cite{Zha15}. The operation is particularly challenging when dealing with large targets such as Envisat \cite{ENVISAT}, for which collision avoidance with solar arrays or antennas during the final phase of rendezvous is critical.

Model Predictive Control (MPC) stands out as a particularly effective technique for this task. MPC is a form of control in which the actuations are computed by solving a finite-time optimization problem at each time step with a receding horizon, using a model of the system to predict its future behavior \cite{Wan09}. The control community over the past decade has channelled significant efforts into refining MPC. As a result, contemporary MPC is backed by a robust theoretical foundation rooted in Lyapunov and invariant set theories, facilitating the development of recursively feasible and intrinsically stable controllers, regardless of whether they rely on linear or nonlinear predictive models \cite{Raw19}. 

The capability of MPC to handle constraints and to optimize control actions in real-time makes it well-suited for complex and safety-critical missions, such as spacecraft rendezvous with cooperative targets. Consequently, the application of Model Predictive Control (MPC) for spacecraft rendezvous \cite{Har12, Leo14} and formation flying \cite{Mor14} has been extensively studied in existing literature. The study by \cite{Ric03} delves into the advantages of both robust and non-robust MPC techniques for rendezvous, contrasting them with other methodologies. The research presented in \cite{How01} considers the impact of velocity measurement errors during formation flight. Furthermore, \cite{Gav12} introduces a robust chance-constrained MPC scheme, aiming to minimize the likelihood of constraint violations in the rendezvous equation influenced by stochastic disturbances. This scheme was later expanded upon for rendezvous in Halo orbits by \cite{San20a} or in  six-degree-of-freedom rendezvous scenarios~\cite{San20b}.

ADR-focused results also exist, primarily centered around 3-DOF rendezvous with non-cooperative tumbling targets. However, in scenarios where constraints vary with time, difficulties might arise when finding a feasible solution for the initial configuration and guaranteeing that it remains feasible for an arbitrary time, requiring very high prediction horizons or a reference optimal trajectory, to the detriment of computational cost and complexity. This issue, known as the 'feasibility problem', does indeed take place in the problem at hand. For instance, \cite{behrendt2023autonomous} solved the time-constrained rendezvous problem for both position and attitude, considering computational limitations. However, collision avoidance was not assessed explicitly. \cite{Liq17} tackled the rendezvous operation assuming attitude synchronization, sidestepping any explicit uncertainty models and resorting to quadratic programs with linear constraints. The considered collision avoidance strategy restricts the position of the chaser outside a sphere around the target, decoupling its dependence on attitude but reducing its applicability to only scenarios in which the docking port coincides with the furthest point from the center of mass of the debris. More intricate strategies have been put forth in works like \cite{Cau5}, which segmented the docking strategy into phases to enhance maneuver safety. When dealing with uncertainties intrinsic to the system, MPC has demonstrated robustness under mild conditions. If a model for these uncertainties is available, it can be integrated into the prediction design, augmenting system performance and bolstering safety. Tube-based approaches have emerged as pivotal in this regard \cite{Limon09}. For instance, \cite{Mam18} uses Tube-Based MPC to perform a rendezvous operation, although no tumbling is considered for the target. Similarly, \cite{Buc18, TubeBased} tackle the rendezvous with tumbling targets using a robust Tube-Based MPC control, computed around a reference trajectory computed by a motion planner. In tandem with the surge of machine learning, enhanced predictive models have been crafted. These models not only predict trajectories but also offer uncertainty estimations from pre-existing data. Some of these models can even update in real-time, leading to the inception of cutting-edge predictive controllers grounded in both offline and online learning methodologies \cite{Man20, Mai21}.  Traditional stabilizing designs of predictive controllers typically focus on steering the system to a fixed, known target equilibrium point, addressing the regulation problem \cite{Raw19}. However, in dynamic environments where the target may evolve or is represented as a trajectory rather than a fixed point, designing stabilizing controls becomes intricate. Recent advancements have yielded predictive controllers proficient in assuring stability and convergence even for moving references \cite{Lim18}. In \cite{Kaikai}, a multistage MPC for Tracking (MPCT) for LTI systems is proposed to achieve safe and stable-by-design 3-DOF rendezvous with tumbling targets, providing recursive feasibility guarantees. Collision avoidance was handled by implementing a more sophisticated and realistic LOS constraint, although only periodic motion of the target was assessed. 

In this work, a novel approach to 3-DOF rendezvous with non-cooperative tumbling targets is introduced. The proposed control algorithm proves to be of high interest for the terminal phases of the rendezvous operation, as it is stable and safe-by-design without assuming simplifications on the target's attitude dynamical model. In particular, a controller based in MPCT for LTV systems is developed through an intermediate coordinate transformation to deal with both the target's three-dimensional non-periodic rotational dynamics and the system's restrictive engineering constraints, namely control, velocity and Line of Sight (LOS). The control is computed as the solution of a Quadratic Programming (QP) problem with linear constraints and a small number of decision variables, that provides feasibility and stability guarantees by means of a terminal LQR and dead-beat region. This approach leads a very low computational cost, ideal for on-line implementation in embedded systems. Note that the designed controller computes the complete control signal, without depending on a possibly costly motion planner. Furthermore, various interesting results are obtained as to determine if a safe docking maneuver can be performed at all given a target rotational state and a control authority for the chaser, serving as useful tools for the decision making process prior to the close-operation. The controller is tested in simulation for realistic conditions, namely, the proximity phase of a rendezvous with Envisat, yielding promising results.

This paper is structured as follows. Section \ref{Sec2} introduces the rendezvous problem, identifying the system's dynamics and constraints, as well as the considered simplifying hypotheses. Section \ref{Sec3} provides firstly a concise overview of the proposed control strategy, followed by a detailed description of state, control, and equilibrium definitions. Section \ref{Sec4} includes the complete design of the controller, including the characterization of terminal virtual control policies, feasibility constraints and weighting matrices. In Section \ref{Sec5}, the implementation and properties of the controller is assessed regarding feasibility, stability, optimality and computational efficiency. The proposed control system is evaluated in simulation in Section \ref{Sec6} for a highly restrictive scenario for a rendezvous mission with Envisat, verifying the practical interest of the proposed approach. Finally, some concluding remarks and future lines of work are described in Section \ref{Sec7}.

\section{Problem formulation}\label{Sec2}

Let $D$ be the target of the rendezvous operation. In this work, it is assumed that $D$ describes a geocentric circular orbit or radius $R$ and angular rate $n = \sqrt{\frac{\mu}{R^3}}$, where $\mu = 398600.4\mathrm{~km}^3 \mathrm{s^{-2}}$ is Earth's planetary constant. The corresponding period is given by $T_D = 2\pi/n$. Note that the extension of the proposed algorithm to elliptical orbits is possible by using the corresponding state transition matrices (see \cite{Yamanaka}). A chaser $C$ spacecraft is to be steered to dock $D$ satisfying all the engineering constraints described later on.

\begin{figure}[htb!]
\centering
\includegraphics[width=0.4\textwidth]{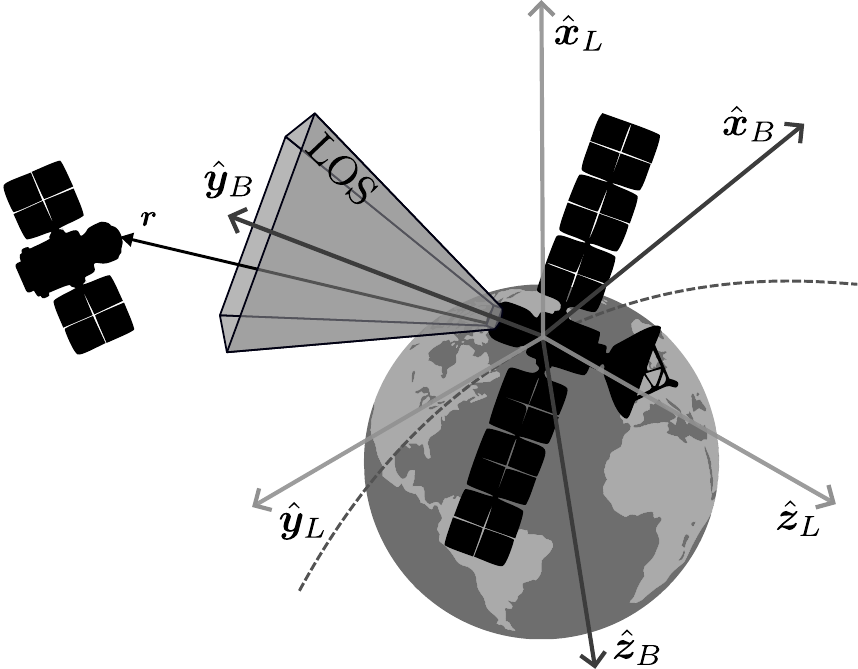}
\caption{Rendezvous operation considering debris body (B) and LVLH (L) reference frames.}
\label{fig:frames}
\end{figure}

Let the chaser to target relative position and velocity be given by vectors $\bm{r}$ and $\bm{v}$. It is convenient to describe these magnitudes in a Local Vertical Local Horizontal (LVLH) frame for the target shown in Figure \ref{fig:frames}, in what follows, given by $L$. Let $\bm{r}_D$ and $\bm{v}_D$ be the position and inertial velocity of the target, and $\bm{h}_D = \bm{r}_D\times\bm{v}_D$ the corresponding orbit angular momentum. The orthonormal basis of $L$ is in general given by
\begin{align}
\hat{\bm{x}}_L = & \frac{\bm{r}_D}{|\bm{r}_D|},\, 
\hat{\bm{y}}_L =  \frac{\bm{h}_D\times\bm{r}_D}{|\bm{h}_D\times\bm{r}_D|} ,\
\hat{\bm{z}}_L =  \frac{\bm{h}_D}{|\bm{h}_D|}.
\end{align}

The local linearization of the relative motion of $C$ and $D$ is given by the Hill-Clohessy-Wiltshire (HCW) equations (see \cite{Hill} and \cite{Clohessy}). In particular, for an impulsive control applied to the chaser $\bm{u}$ and a propagation time $T$, defining the constant quantities $s = \sin{nT}$ and $c = \cos{nT}$, the linearized evolution from instant $t_1 = kT$ to $t_2 = t_1 + T = (k+1)T$ can be written in $L$ as
\begin{equation} \begin{pmatrix} \bm{r}^L(k+1) \\ \bm{v}^L(k+1) \end{pmatrix} = A_{L} \begin{pmatrix} \bm{r}^L(k) \\ \bm{v}^L(k) \end{pmatrix} + B_{L} \bm{u}^L(k). \label{HCW} \end{equation}
where $\bm{r}^L(k)$ denotes the relative position at $t_1 = Tk$ and similarly for other variables. The State Transition Matrices (STM) in \eqref{HCW} are given by
\begin{align}
A_{L} = & {\left[\begin{array}{cccccc}
4-3 c & 0 & 0 & \frac{s}{n} & \frac{2(1-c)}{n} & 0 \\
6(s-n T) & 1 & 0 & -\frac{2(1-c)}{n} & \frac{4 s-3 n T}{n} & 0 \\
0 & 0 & c & 0 & 0 & \frac{s}{n} \\
3 n s & 0 & 0 & c & 2 s & 0 \\
-6 n(1-c) & 0 & 0 & -2 s & 4 c-3 & 0 \\
0 & 0 & -n s & 0 & 0 & c
\end{array}\right] }, \\
B_{L} = & {\left[\begin{array}{c}
0_{3\times 3} \\
I_3
\end{array}\right] }.
\end{align}
The target is modeled as a rigid body that rotates freely around its center of mass. In particular, let $\prescript{B}{L}{C}(k)$ be the rotation matrix from $L$ coordinates to the target's body axes $B$ at time $kT$. Given a vector $\bm{a} = \begin{pmatrix} a_x & a_y & a_z \end{pmatrix}^T \in \mathbb{R}^3$, the cross product matrix operator is defined as
\begin{equation} \bm{a}^\times = \begin{bmatrix} 0 & -a_z & a_y \\ a_z & 0 & -a_x \\ -a_y & a_x & 0 \end{bmatrix}. \end{equation}
For a given initial attitude and angular velocity of the target $\prescript{B}{L}{\bm{\omega}}^B$, the evolution of this matrix is written as the solution of the Euler-Poinsot differential equations, depending on the target's inertia tensor $I_D^B$ in body axes.
\begin{align} \prescript{B}{L}{\dot{\bm{\omega}}}^B = & \left(\prescript{B}{L}{\omega}^B\right)^\times \prescript{L}{E}{\omega}^B -\left(I_D^B\right)^{-1} \left(\prescript{B}{E}{\bm{\omega}}^B\right)^\times I_D^B \prescript{B}{E}{\bm{\omega}}^B \label{EulPoinsot}  \\ \prescript{B}{L}{\dot{C}} = & -\left(\prescript{B}{L}{\bm{\omega}}^B\right)^\times \prescript{B}{L}{C}, \label{CDiffEq} \end{align}
where $\prescript{L}{E}{\omega}^B = n \hat{\bm{z}}_L^B = n \prescript{B}{L}{C} \begin{pmatrix} 0 & 0 & 1 \end{pmatrix}^T$ is the angular velocity of the LVLH frame with respect to the ECI frame $E$ measured in body axes and
\begin{equation} \prescript{B}{E}{\bm{\omega}}^B = \prescript{B}{L}{\bm{\omega}}^B + \prescript{L}{E}{\bm{\omega}}^B. \end{equation}
Note that, if $\left\|\prescript{L}{E}{\bm{\omega}}^B\right\| = n \ll \left\|\prescript{B}{L}{\bm{\omega}}^B\right\|$, equation \eqref{EulPoinsot} can be approximately written as if $L$ was inertial, namely,
\begin{equation} \prescript{B}{L}{\dot{\bm{\omega}}}^{B} \approx -\left(I_D^B\right)^{-1} \prescript{B}{L}{\bm{\omega}}^B \times I_D^B \prescript{B}{L}{\bm{\omega}}^B. \end{equation}
The attitude of the chaser, in the other hand, is assumed to match the one required to perform all maneuvers and therefore the angular dynamics of $C$ are not considered in this work.

The rendezvous maneuver must satisfy several engineering constraints. In order to avoid collisions with the target, the chaser must stay within a safety region, which rotates solidly with $D$, as it given by the target's geometry. Many descriptions of this allowed region have been proposed. For instance, \cite{MotionPlanning} makes use of "capsules" approximating the geometry of both chaser and target to handle collisions. In \cite{OptimalRendezvous}, the minimum distance between the center of mass of chaser and target spacecraft is constrained, generating the safety region as the outside of a sphere centered in the target. In all practical scenarios, a Line of Sight (LOS) prism can be attached to the target's intended rendezvous point. This defines a safe approach region, ensuring collision avoidance with the target. This formulation is convenient for this work, as constraints are convex and linear. In particular, the LOS constraints are time-invariant in $B$, so that
\begin{equation} \begin{bmatrix}
0 & -1 & 0 \\
c_x & -1 & 0 \\
-c_x & -1 & 0 \\
0 & -1 & c_z \\
0 & -1 & -c_z
\end{bmatrix} \bm{r}^B \le \begin{bmatrix} 0 \\ c_x x_0 \\ c_x x_0 \\ c_z z_0 \\ c_z z_0 \end{bmatrix}.
\end{equation}
A representation of the LVLH and body reference frames, together with the relative position and LOS constraint prism, is provided in Figure \ref{fig:frames}.

As for the control vector, it is usually bounded in terms of its components in the chaser's body axes, this is, in a frame rotating in solidity with the actuators. In this scenario, it is assumed that the chaser's attitude control tracks the one of the target, so that these constraints can be written in $B$ axes. In general, depending on the actuators layout, a linear constraint as $A_u \bm{u}^B \le \bm{b}_u$ can be written. Furthermore, it is reasonable to assume that actuators are aligned with the chaser's body axes, leading to
\begin{equation} u_{min} \le u^B_{i} \le u_{max}, i = 1,2,3. \end{equation}
As to guarantee a correct operation of the chaser's on-board systems, and in particular for relative state sensors, additional linear constraints in position and velocity may be written. Therefore, more compactly, the state and control input constraints can be written as
\begin{align} A_x \begin{pmatrix} \bm{r}^B \\ \bm{v}^L \end{pmatrix} \le & b_x \label{Constr_x} \\ A_u \bm{u}^B \le & b_u. \label{Constr_u} \end{align}
A control policy for the terminal phase of rendezvous must ensure these constraints are satisfied at all times. Indeed, any violation of the considered inequalities can be critical to the safety of the whole operation. Furthermore, the controller must be efficient in terms of energy or fuel consumption. For instance, in both chemical and electrical propulsion systems, the total impulse consumed over a time $t = NT$ is related to the fuel mass consumption $\Delta m$ through Tsiolkovsky rocket equation, namely,
\begin{equation} \Delta v = I_{sp}g_0 \ln{\frac{m_0}{m_0 - \Delta m}}, \label{Tsiolkovsky} \end{equation}
where $g_0 = 9.81\mathrm{~m/s^2}$ is the standard gravity, $m_0$ is the initial mass of the spacecraft and $I_{sp}$ is the thruster's specific impulse. Note that a square sum of impulses is given by a quadratic function of the control inputs.
\begin{equation} \sum_{k=1}^{N}{\Delta v(k)^2} = \sum_{k = 1}^{N}{\left\|\bm{u}(k)\right\|}^2. \label{TotalImpulse} \end{equation}
Therefore, it is of interest to both reduce the control effort and the total time $t$ required for the rendezvous operation. The further needs of the implemented control loop to be computationally accessible for spaceborne embedded systems while guaranteeing efficiency and safety motivates the algorithm developed in this work.

\section{MPCT-based rendezvous}\label{Sec3}
In general, the safe rendezvous operation can be written as a constrained optimization problem, the structure of which is given by the considered control framework, and has an effect on the properties and guarantees of the controller. As discussed in Section \ref{Sec5}, the approach considered herein, which based on a novel extension of MPCT for rotation-based LTV systems, leads to a promising results in terms of computational cost, optimality and feasibility. In particular, this strategy leads to the QP problem in \eqref{QP_CostFun}--\eqref{QP_DBEqConstr}, which is solved for each iteration $k$,
\begingroup
\allowdisplaybreaks
\begin{align}
\begin{split}
V&(k,x_k,U(k),\bar{\theta}(k)) = \\&~~ \sum_{i = 0}^{N_p-1}{\left\|x(k+i|k) - x_s(k+i|k)\right\|_Q^2}\\ &+ \sum_{i=0}^{N_p}{\left\|u(k+i|k) - u_s(k+i|k)\right\|_R^2} \\ & + \left\|u(k+N_p|k) - u_s(k+N_p|k)\right\|_R^2 \\&+ \left\|\bar{\theta}(k)\right\|_T^2
\end{split} \\
V^*(k,x_k) & = \min_{U(k),\bar{\theta}(k)} V\left(k,x_k,U(k),\bar{\theta}(k)\right) \label{QP_CostFun} \\
\mathrm{s.t.}&\nonumber \\
U(k) & = \begin{pmatrix} u(k|k) & \hdots & u(k+N_c-1|k) \end{pmatrix} \label{QP_DefU} \\
x(k|k) & = x_k \label{QP_IniConds} \\
\begin{split}
x(k+i+1|k) & = A(k+i) x(k+i|k) + B(k+i) u(k+i|k), \\&~~i = 1, 2, ..., N_p - 1 \label{QP_PropEq}
\end{split}\\
\begin{pmatrix} x_s(k+i|k) \\ u_s(k+i|k) \end{pmatrix} & = M(k+i) \bar{\theta}(k), ~~i = 1, 2, ..., N_p \label{QP_EqPoints} \\
\begin{split}
u(k+i|k) & = K_{LQR}(k+i)\left(x(k+i|k) - x_s(k+i|k)\right) \\&+ u_s(k+i|k),~~i = N_c, ..., N_p - 2 \label{QP_LQR}
\end{split} \\
\begin{split}
u(k+i|k) & = K_{DB,x}(k+i)x(k+i|k) \\&+ K_{DB,\theta}(k+i)\bar{\theta}(k),~~i = N_p-1, N_p \label{QP_DB}
\end{split} \\
A_x x(k+i|k) & \le b_x, A_u u(k+i|k) \le b_u, ~~i = 1, 2, ..., N_p \label{QP_Constr} \\
A_x x_s(k+i|k) & \le b_x, A_u u_s(k+i|k) \le b_u, ~~i = 1, 2, ..., N_p \label{QP_EqConstr} \\
A_{DB} \bar{\theta}(k) & \le b_{DB}. \label{QP_DBEqConstr}
\end{align}
\endgroup
Firstly, the reader is presented with a concise overview of the problem in \eqref{QP_CostFun}--\eqref{QP_DBEqConstr}. Subsequently, a detailed analysis of the definition of variables, linear transformations and constraints is provided.

\subsection{Brief review of the optimization problem}
The optimization problem is formulated in terms of two optimization variables. As shown in \eqref{QP_DefU}, $U(k)$ is defined as the sequence of predicted control inputs starting at $k$ and up to a \textit{control horizon} $N_c$. As usual in MPC, only the fist control input $u(k|k)$ is applied, following a receding horizon fashion. In the framework of MPCT, a new optimization variable $\bar{\theta}(k) \in \mathbb{R}^3$ is introduced, which in what follows is denoted as \textit{equilibrium parameter}. Equation \eqref{QP_CostFun} is given by the target cost function, which is a quadratic function of the system's state $x$ and control input $u$ with respect to a reference \textit{equilibrium trajectory} $\left\{x_s, u_s\right\}$. Note that the cost function also includes a quadratic term depending on the equilibrium parameter. Therefore, the controller steers the system to an equilibrium trajectory, whose distance to the origin is penalized in the cost function. As discussed in detail throughout this work, this added degree of freedom permits the controller to plan its approach based on feasible trajectories rather than a fixed regulation point, extensively improving its behavior in terms of feasibility and region of attraction, while requiring a very low computational effort.

The horizon of the cost function is given by the \textit{prediction horizon} $N_p$, which is in general much greater than the control horizon. Therefore, after the control horizon, explicit control laws introduced in Section \ref{Sec4} and given in \eqref{QP_LQR} and \eqref{QP_DB} are applied, leading to a significant reduction of the number of optimization variables and improving computational performance. The predicted state is initialized at $k$ to the current value, as shown in \eqref{QP_IniConds}, which is assumed to be observable. For a given control policy, the successive values of the state are determined recursively as shown in \eqref{QP_PropEq}. Note that the propagation law is linear but time dependent.

The equilibrium trajectory can be computed as a linear transformation of the equilibrium parameter, as given by \eqref{QP_EqPoints}. Note that, although $\bar{\theta}(k)$ is a constant for each time step $k$, the map $M(k+i)$ is time-dependent, and so is the corresponding trajectory of states and control inputs. The realization of state and control inputs for a candidate solution are required to verify time-independent linear constraints, as written in \eqref{QP_Constr}. These constraints are equally imposed for the equilibrium trajectory, shown in \eqref{QP_EqConstr}. Finally, a set of linear constraints given by \eqref{QP_DBEqConstr} are directly applied to the equilibrium parameter.

\subsection{Definition of state and control variables}
The structure of the optimization problem in \eqref{QP_CostFun}--\eqref{QP_DBEqConstr} is such that all constraints are time-invariant. This is particularly convenient, as it allows to mathematically prove the recursive feasibility of the controller, as detailed later on. Note, however, that the state and control definitions used for propagation in \eqref{HCW}, this is, measured in the LVLH reference frame, do not coincide with the ones for which constraints are constant inequalities, included in \eqref{Constr_x} and \eqref{Constr_u}. For this reason, a partial change of reference frame is introduced. 

If the initial attitude and angular velocity of the target are known, the corresponding rotation matrix evolution $\prescript{B}{L}{C}(k)$ can be computed up to an arbitrary horizon from \eqref{EulPoinsot} and \eqref{CDiffEq}. For computational convenience, it might be of interest to propagate the attitude using a quaternion attitude representation, and generate the sequence $\prescript{B}{L}{C}(k)$ as an output. For the described attitude matrix, the following linear transformations hold for any given vector $\bm{a}$,
\begin{align} \bm{a}^B(k) = & \prescript{B}{L}{C}(k)\bm{a}^L(k) \\ \bm{a}^L(k) = & \left(\prescript{B}{L}{C}(k)\right)^T \bm{a}^B(k). \end{align}
Substituting in \eqref{HCW},
\begin{equation} \begin{aligned} \begin{pmatrix} \left(\prescript{B}{L}{C}(k+1)\right)^T\bm{r}^B(k+1) \\ \bm{v}^L(k+1) \end{pmatrix} = & A_{L} \begin{pmatrix} \left(\prescript{B}{L}{C}(k)\right)^T \bm{r}^B(k) \\ \bm{v}^L(k) \end{pmatrix} \\+ & B_{L} \left(\prescript{B}{L}{C}(k)\right)^T \bm{u}^B(k). \label{HCW_2} \end{aligned} \end{equation}
It is useful to decompose $A_L$ and $B_L$ into the submatrices connecting position, velocity and control, this is,
\begin{align}
A_L = & \begin{bmatrix} A_{L,rr} & A_{L,rv} \\ A_{L,vr} & A_{L,vv} \end{bmatrix} \label{ALDecop} \\
B_L = & \begin{bmatrix} B_{L,ru} \\ B_{L,vu} \end{bmatrix}.
\end{align}
Therefore, and given that $\prescript{B}{L}{C}(k)$ is known beforehand, \eqref{HCW_2} can be written as
\begin{equation}
\begin{pmatrix} \bm{r}^B(k+1) \\ \bm{v}^L(k+1) \end{pmatrix} = A_B(k) \begin{pmatrix} \bm{r}^B(k) \\ \bm{v}^L(k) \end{pmatrix} + B_B(k) \bm{u}^B(k),
\label{EqsOfMotion} \end{equation}
where
\begin{align}
A_B(k) = & \begin{bmatrix} \prescript{B}{L}{C}(k+1) A_{L,rr} \left(\prescript{B}{L}{C}(k)\right)^T & \prescript{B}{L}{C}(k+1) A_{L,rv} \\ A_{L,vr} \left(\prescript{B}{L}{C}(k)\right)^T & A_{L,vv} \end{bmatrix} \label{A_B_Mat} \\
B_B(k) = & \begin{bmatrix} \prescript{B}{L}{C}(k+1) B_{L,ru} \left(\prescript{B}{L}{C}(k)\right)^T \\ B_{L,vu} \left(\prescript{B}{L}{C}(k)\right)^T \end{bmatrix}. \label{B_B_Mat}
\end{align}
For the modified state $x(k) = \begin{pmatrix} \bm{r}^B(k) \\ \bm{v}^L(k) \end{pmatrix}$ and control input $u(k) = \bm{u}^B(k)$, all constraints are time-invariant, directly as written in \eqref{Constr_x}, \eqref{Constr_u}. Therefore, if a position given by the first three components of $x(k)$ is chosen, and provided that it satisfies the LOS constraints at $k$, it is guaranteed to verify them at all future time steps. The cost incurred of introducing this new definition is that the propagation equations are no longer time-invariant. Indeed, the $A(k)$ and $B(k)$ matrices in \eqref{QP_PropEq} are given by $A_B(k)$ and $B_B(k)$, as written in \eqref{A_B_Mat} and \eqref{B_B_Mat}.

\subsection{Equilibrium parameter and equilibrium trajectory}
MPCT makes use of equilibrium points of the linear system as decision variables to increase the region of the attraction of the controller and enhance its stability and feasibility guarantees \cite{FerramoscaAUT09}.  For an LTI linear system such as
\begin{equation} x(k+1) = A x(k) + B u(k) \end{equation}
an \textit{equilibrium point} is given by any pair $\{x_e,u_e\}$ verifying the invariant equation
\begin{equation} x_e = A x_e + B u_e. \label{EqPts0} \end{equation}
In this work, this definition is successfully extended to LTV systems by means of \textit{equilibrium trajectories}. For a time instant $k$, an equilibrium trajectory is given by the pair $\{x_e,u_e\}$ satisfying
\begin{equation} x_e = A(k) x_e + B(k) u_e. \label{EqPtsLTV} \end{equation}
Consequently, let the set of all equilibrium trajectories at time $k$, $\mathcal{Z}(k)$ such that
\begin{equation} \mathcal{Z}(k) = \left\{z = \begin{pmatrix} x_e \\ u_e \end{pmatrix} \in \mathbb{R}^{n_x + n_u} : x_e = A(k) x_e + B(k) u_e \right\}. \end{equation}
The equilibrium trajectories can be mapped by means of the equilibrium parameter $\theta\in\mathbb{R}^{n_\theta(k)}$, for $n_\theta(k)$ the dimension of the subspace of equilibrium points within $\mathbb{R}^{n_x + n_u}$ at time $k$,
\begin{equation} z = M(k)\theta = \begin{bmatrix} M_x(k) \\ M_u(k) \end{bmatrix} \theta, \forall z \in \mathcal{Z}(k). \label{EqPts1} \end{equation}
The dimension $n_\theta(k)$ is in principle time-varying, as are matrices $A(k)$ and $B(k)$. Introducing \eqref{EqPts1} in \eqref{EqPts0}, $M(k)$ can be derived as
\begin{equation} \begin{bmatrix} A(k) - I_{n_x} & B(k) \end{bmatrix} M(k) = 0. \end{equation}
This is equivalent to requiring the columns of $M(k)$ to be a linear basis of the kernel of $\begin{bmatrix} A(k) - I_{n_x} & B(k) \end{bmatrix}$, this is,
\begin{equation} M(k) = \ker{\begin{bmatrix} A(k) - I_{n_x} & B(k) \end{bmatrix}}. \label{MDef} \end{equation}
The following result, proven in Appendix \ref{Ap1}, allows to define a consistent framework of equilibrium trajectories for the problem at hand.
\begin{lemma}
The dimension of the space of equilibrium points for the LTV system given in \eqref{EqsOfMotion} is constant and equal to $n_\theta(k) = 3$. Furthermore, it is possible to define $M(k)$ such that $\theta$ is equal to the position of the considered equilibrium point in $B$, $\theta = \bm{r}^B_e$, which is constant by definition.
\end{lemma}
The most useful consequence of these two results is that $M(k)$ can be written, by performing the corresponding linear combination of columns, in the following triangular form,
\begin{equation} M(k) = \begin{bmatrix} 1 & 0 & 0 \\ 0 & 1 & 0 \\ 0 & 0 & 1 \\ \vdots & \vdots & \vdots \end{bmatrix}. \label{MStruct} \end{equation}
In this formulation, the bijection $\theta\leftrightarrow\bm{r}^B_e(k)$ is trivially the identity, which means that equilibrium trajectories are uniquely associated to their constant position. Hence, if an equilibrium trajectory associated to the equilibrium parameter $\theta$ initially verifies the LOS constraints, it does verify them for all times. Additionally, the computation of \eqref{MStruct} is straightforward and computationally efficient. For this purpose, any basis of the kernel in \eqref{MDef} can be obtained, and then be written in column echelon form to derive $M(k)$ as defined in this work.\footnote{In computational algebra environments, this operation can typically be obtained by transposing the initial basis, computing its row echelon form and transposing the result.}

\section{Design of the controller}\label{Sec4}
Once state, control and equilibrium trajectories are properly defined, it remains to characterize the virtual controllers in \eqref{QP_LQR} and \eqref{QP_DB} and derive the additional equilibrium parameter constraint in \eqref{QP_DBEqConstr}. These ingredients are closely tied to the properties of the controller, for instance, the controller in \eqref{QP_LQR} is implemented to enhance optimality and stability, while both the controller in \eqref{QP_DB} and the constraint \eqref{QP_DBEqConstr} guarantee recursive feasibility. Furthermore, the controller requires the definition of several weight matrices, for both the cost function \eqref{QP_CostFun} and the terminal LQR \eqref{QP_LQR}. In this section, all these topics are covered in detail, leading to a complete characterization of the controller.

\subsection{Terminal controllers design}
\begin{figure}[htb!]
\centering
\includegraphics[width=0.4\textwidth]{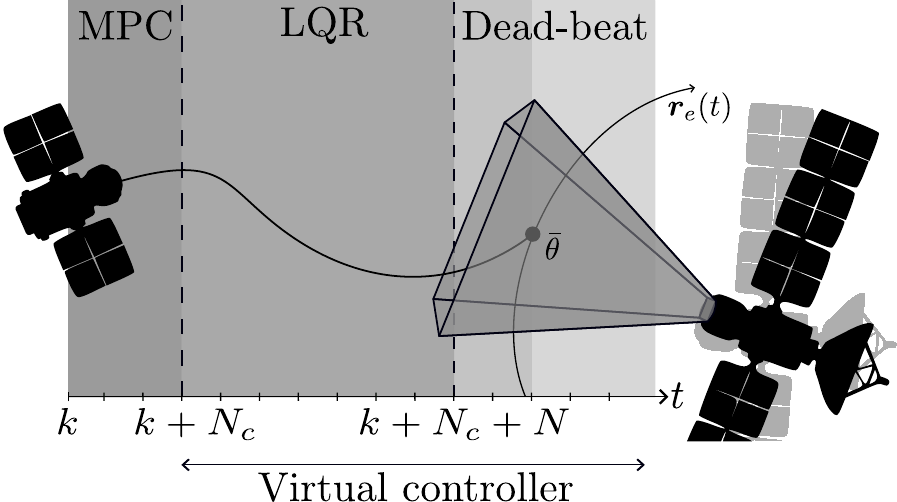}
\caption{Real and virtual controller layout with feasibility and stability guarantees.}
\label{fig:controller}
\end{figure}

After the control horizon, the optimizer explicitly computes a control input given by a two-phases policy. This leads to a terminal \textit{virtual controller} applied up to the prediction horizon, as displayed schematically in Figure \ref{fig:controller}.
In broad outline, between $k+N_c$ and $k+N_p-2 = k + N_c+N$, an infinite-time time-varying LQR, derived from a rotation of the LVLH frame, is used to steer the state towards the target equilibrium trajectory. At time-instants $k+N_p-1$ and $k+N_p$, an explicit dead-beat controller is used to reach the target equilibrium trajectory from the final state after the LQR application. Finally, from $k+N_p+1$ onwards, a dead-beat controller is used to track exactly the target equilibrium trajectory.

An infinite-time time-varying LQR for a moving target can be derived by constructing the controller in the LVLH reference frame, in which dynamics are LTI. Indeed,
\begin{equation} \begin{split} \bm{u}^L(k+i|k) & = K_{F} \left(\begin{pmatrix} \bm{r}^L(k+i|k) \\ \bm{v}^L(k+i|k) \end{pmatrix} - \begin{pmatrix} \bm{r}^L_s(k+i|k) \\ \bm{v}^L_s(k+i|k) \end{pmatrix} \right) \\&+ \bm{u}_s^L(k+i|k). \end{split} \end{equation}
The constant gain $K_F$ is directly obtained as
\begin{equation} K_{F} = -\left(R_{LQR} + B_L^T P_{LQR} B_L\right)^{-1} B_L^T P_{LQR} A_L. \end{equation}
The Lyapunov cost $P_{LQR}$ is computed by solving the algebraic Riccati equation in LVLH axes, considering the weighting matrices $Q_{LQR}$ and $R_{LQR}$ for state and control, respectively, so that
\begin{equation} P_{LQR} = Q_{LQR} + A_L^T P_{LQR} A_L + A_L^T P_{LQR} B_L K_{LQR}. \end{equation}
Thus, expressing the control law in the modified reference frame,
\begin{equation} \begin{split}
u(k+i|k) & = K_{LQR}(k+i)\left(x(k+i|k) - x_s(k+i|k)\right) \\&+ u_s(k+i|k)
\end{split} \end{equation}
where
\begin{equation} K_{LQR}(k+i) = \prescript{B}{L}{C}(k+i) K_{F} \begin{bmatrix} \prescript{B}{L}{C}(k+i)^T & 0_{3} \\ 0_{3} & I_{3} \end{bmatrix}. \end{equation}
The weighting matrices $Q_{LQR}$ and $R_{LQR}$ do not necessarily need to coincide with the ones used in the MPC cost function, although that configuration might be a good initial fix before tuning. Note that $K_{LQR}(k+i)$ does not depend on the state nor the control input, this is, can be computed prior to the optimization step for each $k$. Consequently, the LQR control in \eqref{QP_LQR} is indeed explicit for each time instant. This allows considering long horizons $N$ for the application of the LQR without affecting computational performance, since the dimension of the optimization problem is unaltered.

Immediately after the application of the LQR, a two-iterations dead-beat controller is implemented to reach the target equilibrium trajectory. For the sake of clarity, the development of this controller stage is first computed in the LVLH frame and then rotated to match the structure in \eqref{QP_DB}. Without accounting for constraints, a two-steps explicit controller can steer the system from any state to a target equilibrium trajectory, as the matrix $\begin{pmatrix} B_L & A_L B_L \end{pmatrix}$ is full rank. Furthermore, due to the structure of $B_L$, a control action $\bm{u}^L(k)$ does only have an effect on the state at iteration $k+2$. Hence, taking $\bm{u}^L(k+N_p-1|k)$ as the only input that steers $\bm{r}^L(k+N_p+1|k)$ to the position of the target equilibrium trajectory, $\bm{r}_e^L(k+N_p+1|k)$, and $\bm{u}^L(k+N_p|k)$ such that $\bm{r}^L(k+N_p+2|k) = \bm{r}_e^L(k+N_p+2|k)$, the equilibrium trajectory is reached at $k+N_p+1$ and can be tracked exactly by arbitrarily repeating this procedure.

Let the decomposition of $A_L$ given in \eqref{ALDecop}. The control action $\bm{u}^L(k+N_p-1)$ is given by
\begin{equation} \begin{split} &\bm{u}^L(k+N_p-1|k) \\&= A_{L,rv}^{-1} \bm{r}^L_e(k+N_p+1|k) \\ &- A_{L,rv}^{-1} \left( A_{L,rr}^2 + A_{L,rv} A_{L,vr} \right) \bm{r}^L(k+N_p-1|k) \\&- \left( A_{L,rr} A_{L,rv} + A_{L,rv} A_{L,vv} \right) \bm{v}^L(k+N_p-1|k). \end{split} \end{equation}
Condensing notation,
\begin{equation}\begin{split} u(k+N_p-1|k) & = A_{L,rv}^{-1} \bm{r}^L_e(k+N_p+1|k) \\&+ \hat{K}_{DB} \begin{pmatrix} \bm{r}^L(k+N_p-1|k) \\ \bm{v}^L(k+N_p-1|k) \end{pmatrix}. \end{split} \end{equation}
where
\begin{align} \hat{K}_{DB} & = \begin{pmatrix} \hat{K}_{DB,r} & \hat{K}_{DB,v} \end{pmatrix}\\
\hat{K}_{DB,r} & = - A_{L,rv}^{-1} \left( A_{L,rr}^2 + A_{L,rv} A_{L,vr} \right) \\
\hat{K}_{DB,v} & = - \left( A_{L,rr} A_{L,rv} + A_{L,rv} A_{L,vv} \right). \end{align}
Similarly, for iteration $k+N_p$,
\begin{equation} \begin{split} u(k+N_p|k) & = A_{L,rv}^{-1} \bm{r}^L_e(k+N_p+2|k) \\&+ \hat{K}_{DB} \begin{pmatrix} \bm{r}^L(k+N_p|k) \\ \bm{v}^L(k+N_p|k) \end{pmatrix}. \end{split} \end{equation}
The dependence of the position of the equilibrium point in LVLH axes and $\theta$ is explicitly given for any iteration by
\begin{equation} \bm{r}_e^L(k) = \prescript{B}{L}{C}(k)^T \theta. \end{equation}
Similarly, for the state in LVLH and in the modified reference frame,
\begin{equation} \begin{pmatrix} \bm{r}^L(k) \\ \bm{v}^L(k) \end{pmatrix} = \begin{bmatrix} \prescript{B}{L}{C}(k)^T & 0_3 \\ 0_3 & I_3 \end{bmatrix} x(k). \end{equation}
Consequently, the control actions at $k+i$, with $i = N_u+N$ and $i = N_u+N+1$, are linear functions of $x(N_u+N)$ and $\bar{\theta}$ only,
\begin{equation} u(k+i|k) = K_{DB,x}(k+i) x(k+i|k) + K_{DB,\theta}(k+i) \bar{\theta}(k), \end{equation}
where
\begin{align}
K_{DB,x}(k+i) & = \prescript{B}{L}{C}(k+i) \hat{K}_{DB} \begin{bmatrix} \prescript{B}{L}{C}(k+i)^T & 0_3 \\ 0_3 & I_3 \end{bmatrix} \\
K_{DB,\theta}(k+i) & = \prescript{B}{L}{C}(k+i) A_{L,rv}^{-1} \prescript{B}{L}{C}(k+i+2)^T.
\end{align}

\subsection{Feasibility constraint derivation}
Once the dead-beat controller is applied, the system reaches exactly the target equilibrium point. A second dead-beat is then implemented to track the equilibrium trajectory. In this case, however, a different strategy is considered. Instead of explicitly applying the control constraints to each control input derived from this policy, the maximum required control authority to track a specific equilibrium trajectory up to infinite horizon is computed. This value is then compared to the maximum available control region to determine if the chosen trajectory is feasible. As proven later in this section, the feasibility of an equilibrium trajectory can be required by introducing a single additional linear constraint on the equilibrium parameter $\theta$, as written in \eqref{QP_DBEqConstr}. Furthermore, as proven in Section \ref{Sec5}, the satisfaction of this additional constraint guarantees recursive feasibility. Consequently, this section devotes to finding an explicit expression for the feasibility constraint \eqref{QP_DBEqConstr} by means of a dead-beat controller.

For this analysis, the following results can be considered. Firstly, the position of a given equilibrium point at any time instant remains at a constant distance from the target, this is, $\bm{r}_e^L(k) \in \partial B(\|\theta\|)$, being $B(\|\theta\|)$ a 3-ball of radius $\theta$ and $\partial B(\|\theta\|)$ its contour. Similarly, the norm of the velocity vector can be bounded by the conservation of angular momentum. In particular, $\|\bm{v}_e^L(k)\| \in \left[0, \frac{\|\bm{h}||}{I_3}\right]$, where $\bm{h}$ is the target's angular momentum and $I_3$ is it the inertia moment of the target's minor axis. Furthermore, the velocity of an equilibrium point in LVLH axes is given by $\bm{v}_e^L(k) = \prescript{B}{L}{\bm{\omega}}(k)\times\bm{r}_e^L(k)$, and the evolution of $\prescript{B}{L}{\bm{\omega}}(k)$ is obtained from the Euler-Poinsot equations in \eqref{EulPoinsot}.

\begin{figure}[htb!]
\centering
\includegraphics[width=0.4\textwidth]{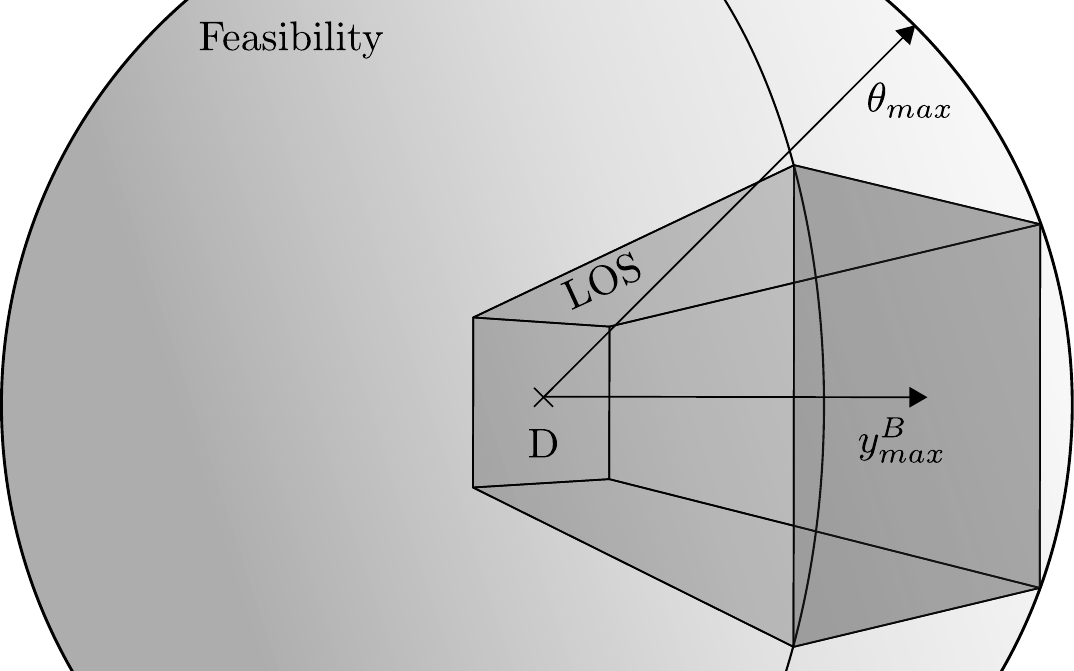}
\caption{Modified LOS region considering feasibility constraints.}
\label{fig:feasibility}
\end{figure}

For small angular velocities with respect to the sampling time $T$, the motion of a given equilibrium point can be written as
\begin{equation} \bm{r}^L_e(k+1) = \bm{r}^L_e(k) + \bm{v}^L_e(k) T. \end{equation}
The position of the chaser is propagated as
\begin{equation} \bm{r}^L (k+1) = A_{L,rr} \bm{r}^L (k) + A_{L,rv} \bm{v}^L (k). \end{equation}
Therefore, the chaser can follow an equilibrium point position only if
\begin{equation} \bm{v}^L (k) = A_{L,rv}^{-1} \left[ \left(I_3 - A_{L,rr}\right) \bm{r}^L_e(k) + \bm{v}^L_e(k) T \right]. \end{equation}
The control itself is involved in the propagation of the chaser's velocity, so that
\begin{equation} \begin{split} \bm{u}^L(k) &= \bm{v}^L(k+1) - A_{L,vv} \bm{v}^L(k) - A_{L,vr} \bm{r}_e^L(k)  \\
&= A_{L,rv}^{-1} \left[ \left(I_3 - A_{L,rr}\right) \bm{r}^L_e(k+1) + \bm{v}^L_e(k+1) T \right] \\ &-A_{L,vv} A_{L,rv}^{-1} \left[ \left(I_3 - A_{L,rr}\right) \bm{r}^L_e(k) + \bm{v}^L_e(k) T \right] \\ &-A_{L,vr} \bm{r}_e^L(k).
\end{split} \end{equation}
The kinematics of the equilibrium points relate position in two consecutive iterations. As for velocity, the following can be written for a small sampling time,
\begin{equation} \begin{split} \bm{v}_e^L(k+1) &= \prescript{B}{L}{\bm{\omega}}^{L}(k+1) \times \bm{r}_e^L(k+1) \\ &= \left(\prescript{B}{L}{\bm{\omega}}^{L}(k) + T \prescript{B}{L}{\dot{\bm{\omega}}}^{L}(k) \right) \times \left(\bm{r}_e^L(k) + \bm{v}_e^L(k) T \right) \\ &= \left(\prescript{B}{L}{\bm{\omega}}^{L}(k) + T \prescript{B}{L}{\dot{\bm{\omega}}}^{L}(k) \right) \\ &~~~\times \left(\bm{r}_e^L(k) + \prescript{B}{L}{\bm{\omega}}^{L}(k) \times \bm{r}_e^L(k) T \right).  \end{split} \end{equation}
Simplifying from orthogonality and retaining terms up to the first order,
\begin{equation} \bm{v}_e^L(k+1) = \bm{v}_e^L(k) + T \prescript{B}{L}{\dot{\bm{\omega}}}^{L}(k) \times \bm{r}_e^L(k) = \bm{v}_e^L(k) + \Delta \bm{v}_e^L(k), \end{equation}
where
\begin{equation} \Delta \bm{v}_e^L(k) = T \prescript{B}{L}{\dot{\bm{\omega}}}^{L}(k) \times \bm{r}_e^L(k). \end{equation}
From the previous results, introducing the first order kinematics in the control law,
\begin{equation} \begin{split} \bm{u}^L(k) &= \left[A_{L,rv}^{-1}A'_{L,rr} - A_{L,vv} A_{L,rv}^{-1} A'_{L,rr}  - A_{L,vr} \right] \bm{r}_e^L(k) \\ 
&+ \left[T A_{L,rv}^{-1} A'_{L,rr} - T A_{L,vv} A_{L,rv}^{-1} + T A_{L,rv}^{-1} \right] \bm{v}_e^L(k) \\
&+ \left[T A_{L,rv}^{-1}\right] \Delta\bm{v}_e^L(k).
\end{split} \end{equation}
where
\begin{equation} A'_{L,rr} = \left(I_3 - A_{L,rr}\right). \end{equation}
In condensed notation, the control is given by the considered equilibrium point as
\begin{equation} \bm{u}^L(k) = \mathcal{U}^L \begin{pmatrix} \bm{r}_e^L(k) \\ \bm{v}_e^L(k) \\ \Delta\bm{v}_e^L(k) \end{pmatrix}, \end{equation}
where $\mathcal{U}^L = \begin{bmatrix} \mathcal{U}^L_r & \mathcal{U}^L_v & \mathcal{U}^L_{\Delta v} \end{bmatrix}$ is a constant matrix. In this regard, $\mathcal{U}^L$ is an invariant matrix for all problems, given a sampling time, for varying rotational states of the target. This control policy can be rotated to body axes $B$, leading to
\begin{equation} \bm{u}^B(k) = \prescript{B}{L}{C}(k) \mathcal{U}^L \left(\prescript{B}{L}{\bar{C}}(k)\right)^T \begin{pmatrix} \bm{r}_e^B(k) \\ \bm{v}_e^B(k) \\ \Delta\bm{v}_e^B(k) \end{pmatrix}, \end{equation}
where
\begin{equation} \left(\prescript{B}{L}{\bar{C}}(k)\right)^T = \begin{bmatrix} \left(\prescript{B}{L}{C}(k)\right)^T & 0_3 & 0_3 \\ 0_3 & \left(\prescript{B}{L}{C}(k)\right)^T & 0_3 \\ 0_3 & 0_3 & \left(\prescript{B}{L}{C}(k)\right)^T. \end{bmatrix} \end{equation}
The proposed control action must satisfy the corresponding constraints, this is,
\begin{equation} A_u \mathcal{U}^B(k) \begin{pmatrix} \bm{r}_e^B(k) \\ \bm{v}_e^B(k) \\ \Delta\bm{v}_e^B(k) \end{pmatrix} \le b_u, \forall k, \label{FinalPolytope} \end{equation}
where
\begin{equation} \mathcal{U}^B(k) = \prescript{B}{L}{C}(k) \mathcal{U}^L \left(\prescript{B}{L}{\bar{C}}(k)\right)^T \end{equation}
Since both $\bm{r}_e^L(k)$ and $\bm{v}_e^L(k)$ are contained in sets with spherical symmetry, and as $\bm{r}_e^B(k)$ and $\bm{v}_e^B(k)$ are obtained by means of a rotation, these vectors too are contained in the aforementioned sets. $\Delta \bm{v}_e^B(k)$ can be similarly characterized by considering the Euler-Poinsot equations. Indeed, neglecting the angular velocity of the LVLH axes with respect to inertial axes, the time derivative of the angular velocity is given by
\begin{equation} \prescript{B}{L}{\dot{\bm{\omega}}}^{B}(k) = -\left(I_D^B\right)^{-1} \prescript{B}{L}{\bm{\omega}}^B(k) \times I_D^B \prescript{B}{L}{\bm{\omega}}^B(k). \end{equation}
Since in general there is no correlation between $\bm{r}_e^B(k) = \theta$ and $\prescript{B}{L}{\dot{\bm{\omega}}}^B(k)$, the norm of the velocity increment must be bounded as
\begin{equation}  \|\Delta \bm{v}_e^B(k)\| \le T\left\| \left(I_D^B\right)^{-1} \prescript{B}{L}{\bm{\omega}}^B(k) \times I_D^B \prescript{B}{L}{\bm{\omega}}^B(k)  \right\| \|\theta\|. \end{equation}
Solutions for the angular velocity of a rigid body in free rotation, written in body axes, are given by closed curves denoted as polhodes, which can be computed as the intersection of two ellipsoids, namely
\begin{equation} \begin{cases}
\dfrac{\left(\prescript{B}{L}{\omega}^B_1\right)^2}{h^2/I_1^2} + \dfrac{\left(\prescript{B}{L}{\omega}^B_2\right)^2}{h^2/I_2^2} + \dfrac{\left(\prescript{B}{L}{\omega}^B_3\right)^2}{h^2/I_3^2} = 1 \\
\dfrac{\left(\prescript{B}{L}{\omega}^B_1\right)^2}{2E/I_1} + \dfrac{\left(\prescript{B}{L}{\omega}^B_2\right)^2}{2E/I_2} + \dfrac{\left(\prescript{B}{L}{\omega}^B_3\right)^2}{2E/I_3} = 1,
\end{cases} \end{equation}
where $h$ and $E$ are the angular momentum and rotational kinetic energy, respectively. Note that $I_1, I_2, I_3$ are the components of the inertia tensor in body axes. Therefore,
\begin{equation} \|\Delta \bm{v}_e^B(k)\| \le \sqrt{\varpi^*} T\|\theta\|, \end{equation}
where
\begin{equation} \begin{split} \varpi^* = &\max_{\bm{\omega}}{\left\{ \left\|\begin{pmatrix} \omega_1/I_1 \\ \omega_2/I_2 \\ \omega_3/I_3 \end{pmatrix} \times \begin{pmatrix} \omega_1 I_1 \\ \omega_2 I_2 \\ \omega_3 I_3 \end{pmatrix}\right\|^2 \right\}} \\ s.t.& \begin{cases}
\dfrac{\omega_1^2}{h^2/I_1^2} + \dfrac{\omega_2^2}{h^2/I_2^2} + \dfrac{\omega_3^2}{h^2/I_3^2} = 1 \\
\dfrac{\omega_1^2}{2E/I_1} + \dfrac{\omega_2^2}{2E/I_2} + \dfrac{\omega_3^2}{2E/I_3} = 1.
\end{cases}  \end{split} \end{equation}
This problem can be solved explicitly by using Lagrange multipliers, yielding the following expression,
\begin{equation} \varpi^* = \max_{i \ne j \in \{1,2,3\}}{\left\{-\frac{\left(I_i + I_j\right)^2}{\left(I_i I_j\right)^3} \left(h^2 - 2 I_i E\right)\left(h^2 - 2 I_j E\right) \right\}}. \end{equation}
Therefore, from this analysis, without needing to propagate the state of the considered equilibrium point, a constraint can be imposed in $\theta$ by writing \eqref{FinalPolytope} for the bounded sets of $\bm{r}_e^B$, $\bm{v}_e^B$ and $\Delta \bm{v}_e^B$,
\begin{equation} \begin{split} &A_u \mathcal{U}^B(k) \begin{pmatrix} \bm{r} \\ \bm{v} \\ \Delta\bm{v} \end{pmatrix} \le b_u, \\ &\forall~\bm{r} \in \partial B(\|\theta\|), \bm{v} \in B\left(\|\theta\| \frac{\|\bm{h}\|}{I_3} \right), \Delta \bm{v} \in B\left(\sqrt{\varpi^*} T\|\theta\| \right). \end{split} \end{equation}
This time-dependent inequality can be bounded by one that is invariant. For this purpose, let the \textit{matrix 2-norm} operator be given by the following for square matrices
\begin{equation} \|A\| = \sup_{\bm{x}\ne0}{\frac{\|A\bm{x}\|}{\|\bm{x}\|}} = \max_{\lambda\in\sigma(A)}{|\lambda|}. \end{equation}
Adopting a simplified notation for the sake of clarity, let
\begin{equation}  \left\|\mathcal{U}^B(k)\right\| = \left\|\mathcal{U}^L\right\| = \begin{bmatrix} I_3\|\mathcal{U}^L_{r}\| & I_3\|\mathcal{U}^L_{v}\| & I_3\|\mathcal{U}^L_{\Delta v}\| \end{bmatrix}. \end{equation}
Taking $\bm{r}$, $\bm{v}$ and $\Delta \bm{v}$ within their corresponding spherically symmetric sets, it follows that
\begin{equation} \max_{\bm{r}, \bm{v}, \Delta\bm{v}}{\mathcal{U}^B(k) \begin{pmatrix} \bm{r} \\ \bm{v} \\ \Delta\bm{v} \end{pmatrix}} \le \left\|\mathcal{U}^L\right\| \begin{pmatrix} \bm{r} \\ \bm{v} \\ \Delta\bm{v} \end{pmatrix}. \end{equation}
Since $\partial B(\|\theta\|) \in B(\|\theta\|)$, the constraint can be rewritten only in terms of balls, leading to
\begin{equation} \begin{split} &A_u \left\|\mathcal{U}^L\right\| \begin{pmatrix} x_1 \\ x_2 \\ x_3 \end{pmatrix} \le b_u, \\ &\forall~x_1 \in B(\|\theta\|), x_2 \in B\left(\|\theta\| \frac{\|\bm{h}\|}{I_3} \right), x_3 \in B\left(\sqrt{\varpi^*} T\|\theta\| \right). \end{split}
\label{FinalConstr} \end{equation}
Interestingly, \eqref{FinalConstr} depends solely on the norm of $\theta$, being time invariant. Note that, as the set of feasible vectors $\theta$ is constant, it can be completely computed offline and only once, to be used in all designed controllers. Furthermore, note that the corresponding set is exactly given by a sphere. In other words, this final constraint defines a maximum equilibrium points distance for which the controller is feasible and stable. This condition depends solely on the factors $\|\bm{h}\|/I_3$ and $\sqrt{\varpi^*} T$ as parameters containing the varying information for the target. In consequence, for a given chaser sampling time, the maximum allowable $\theta$ can be computed as a function of the rotational state of the target, completely defining the most critical admissible rendezvous missions for a given technology, and serving as a powerful tool for the chaser design process. Moreover, the maximum allowable value of $\theta$ can be explicitly computed, reducing extensively the needed computational resources to implement the controller. For this purpose, the following result is introduced, as proved in Appendix \ref{Ap1}.
\begin{lemma}
Let the nonzero matrices $A^1, ... , A^l \in \mathbb{R}^{m\times n}$. Let $b\in \mathbb{R}^{m}$, $r\in \mathbb{R}^l$ such that $b\ge0, r>0$. For the scalar $\theta\in\mathbb{R}^+$, let the sets given by n-balls of radius $r_j \theta$, $\Omega_j = B^n(r_j\theta)$, $j = 1, ... , l$. Consider the following inequality,
\begin{equation} \sum_{j=1}^{l}{A^j x_j} \le b, \forall x_j \in \Omega_j. \label{BallsIneq} \end{equation}
The maximum value $\theta$ for which \eqref{BallsIneq} holds is given by
\begin{equation}
\theta_{max} = \min_{i \in \mathbb{N}\cap[1,m]}{\left\{ \frac{b_i}{\sum_{j=1}^{l}{r_j\sqrt{\sum_{k=1}^{n}{\left(A^j_{ik}\right)^2}}}} \right\}}.
\end{equation}
\end{lemma}
This allows to solve \eqref{FinalConstr} explicitly with virtually no associated cost. After computing $\theta_{max}$ offline, the optimization problem can be extended to guarantee stability and feasibility by including the additional constraint on $\theta$, namely,
\begin{equation} \theta^T \theta \le \theta_{max}^2. \end{equation}
The optimization problem is therefore a convex Quadratically Constrained Quadratic Programming (QCQP). An interior approximation of the constraint on $\|\theta\|$ given by a politope can be considered to express the MPC as the solution of a simpler QP problem with linear constraints, which is undoubtedly more favorable from an implementation standpoint. Conveniently, a single polytope can be considered as to the LOS constraint together with the maximum norm for $\theta$, leading to a good approximation with only some additional linear inequalities. In particular, the limit sphere on $\theta$ can be approximated as an $XZ$ plane in body axes, bounding $y^B$ by a maximum threshold, as represented in Figure \ref{fig:feasibility}, given by
\begin{equation} y_{max}^B = c_x c_z \frac{\left( \sqrt{\xi_3  \theta_{max}^2 - \xi_1}-\xi_2 \right) }{\xi_3} \end{equation}
where
\begin{align}
\xi_1 & = c_x^2 c_z^2 (x_0^2 + z_0^2) + (c_z z_0 + c_x x_0)^2 \\
\xi_2 & = c_x z_0 + c_z x_0 \\
\xi_3 &= c_x^2 c_z^2 + c_x^2 + c_z^2.
\end{align}
In consequence, the additional constraint \eqref{QP_DBEqConstr} for the equilibrium parameter is simply given by
\begin{equation} \begin{bmatrix} 0 & 1 & 0 \end{bmatrix} \bar{\theta} \le y_{max}^B. \end{equation}

The extensive analysis performed on the behavior of the target for arbitrary horizon yields to very powerful results. Indeed, for the rendezvous conditions considered in \cite{TubeBased}, an impressive result of $\theta_{max} = 3.597\mathrm{~m}$ is obtained.\footnote{Note that a poorer bounding of $\|\Delta \bm{v}_e^L\| \le T \frac{h^2}{I_3^2} \|\theta\|$, which is computed as a worst-case-scenario without even considering the target dynamics, yields $\theta_{max} = 0.1751\mathrm{~m}$. Moreover, the value obtained by neglecting the dynamics of the target and considering that velocities in $k$ and $k+1$ are independent gives a much worst estimate of $\theta_{max} = 0.05739\mathrm{~m}$. This certainly justifies the added mathematical technicalities, which allow a much tighter bounding.} Since the initial conditions are within this region, the control is guaranteed to be feasible and stable, in all cases. Moreover, $\theta_{max}$ is not the maximum initial distance from the target, but the maximum distance at which the target equilibrium point $\theta$ can be placed by the optimization algorithm for a given iteration. Therefore, a much bigger region of attraction can be expected for the proposed MPC algorithm, depending on the control horizon.

\subsection{Definition of weighting matrices}
In order to completely characterize the controller, several weighting matrices are to be defined. Firstly, the MPC cost function in \eqref{QP_CostFun} is given by $Q$, $R$ and $T$. Secondly, the cost function in the virtual LQR requires values for $Q_{LQR}$ and $R_{LQR}$. As usual, these matrices must be positive definite. For the application at hand, it is convenient to define all weighting matrices as diagonal, allowing for simple comparison as given by the relative sizes of their non-zero terms.

The selection of matrices $Q$, $R$ and $T$ does not affect the feasibility or stability of the controller, as the proposed solution must satisfy all constraints independently of the weighting matrices. Indeed, the definition of the cost function does only affect the prioritization among already feasible solutions. Furthermore, the weighting of the equilibrium parameter, given by $T$, affects how tightly the optimizer places the feasible equilibrium trajectory to the actual target. For a sufficiently large $T$, the optimizer would set $\bar{\theta}=0$ if possible, or to the closest feasible equilibrium trajectory otherwise. This is an interesting property of MPCT for the current application, as using large $T$ enhances optimality properties while retaining feasibility guarantees. Indeed, taking $\bar{\theta}\approx0$ leads the cost function \eqref{QP_CostFun} to have the same structure of a conventional MPC controller with a terminal LQR constraint, which is known to be optimal. Therefore, this strategy was followed in this work. The relative size of $Q$ and $R$ can be tuned following conventional MPC strategies. Typically, position and velocity terms in $Q$ can be weighted with equal values, respectively. Furthermore, $R$ can be written as a scaled identity matrix if the actuator layout is symmetric. Note that the first and last diagonal term in $T$ are associated with non-radial relative position coordinates. Therefore, taking a higher value for these coefficients penalizes equilibrium trajectories that are closer to the walls of the LOS pyramid, allowing for a more robust implementation.

The LQR weighting matrices $Q_{LQR}$ and $R_{LQR}$ can be tuned following similar rules. However, their values do affect feasibility. For instance, if $Q_{LQR} \gg R_{LQR}$, the controller prioritizes approaching the target equilibrium trajectory rapidly, possibly failing to comply with control constraints. Conversely, if $Q_{LQR} \ll R_{LQR}$, control effort is carefully managed, which might lead to breaking the LOS constraint or not approaching sufficiently the target trajectory, with a noncompliance of the control authority limit in the two-step dead-beat phase. It is convenient to use equal terms for all position, velocity and control weights, respectively. Additionally, as the behavior of an LQR is unaffected by scaling factors in its cost function, let $R_{LQR} = I_3$, leading to
\begin{align} Q_{LQR} & = q_{LQR}\begin{bmatrix} I_3 & 0_3 \\ 0_3 & \alpha_{LQR} I_3 \end{bmatrix} \\
R_{LQR} & = I_3.
\end{align}
Hence, only 2 positive scalar values $q_{LQR}$ and $\alpha_{LQR}$ are required to fully characterize the LQR. These coefficient can be tuned, for instance, by means of a Monte Carlo analysis, a simplified dynamical model or pure heuristics.

\section{Implementation and properties of the controller}\label{Sec5}
In Section \ref{Sec4}, a set of sequential virtual controllers was designed to predict the future behavior of the system. In particular, an LQR was developed to optimally approach the equilibrium trajectory. Additionally, a two-step dead-beat controller was constructed to exactly reach the target equilibrium trajectory. Finally, an explicit linear constraint was developed through an auxiliary dead-beat controller to determine the feasibility of an equilibrium trajectory solely in terms of its equilibrium parameter $\bar{\theta}$. This strategy leads to a locally optimal controller, this is, which provides optimal control as long as $\bar{\theta}=0$ is a feasible target state. In this section, it is proven that the aforementioned controller is stable, recursively feasible to infinite horizon and highly computationally efficient.

\begin{figure}[htb!]
\centering
\includegraphics[width=0.45\textwidth]{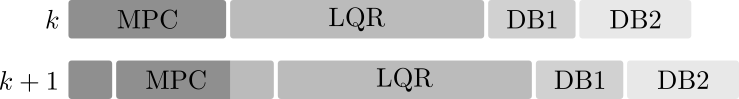}
\caption{Construction of a feasible solution at $k+1$ given a solution at $k$.}
\label{fig:Feas}
\end{figure}

For this purpose, let the optimization problem in \eqref{QP_CostFun}--\eqref{QP_DBEqConstr} at some time instant $k$. Let $U_k$ be a feasible solution of \eqref{QP_CostFun}--\eqref{QP_DBEqConstr} at $k$. Due to the structure of the proposed virtual controller, $U_k$ predicts an infinite sequence of control inputs such as
\begin{equation} U_k = \begin{pmatrix} U_{MPC}(k|k+N_c-1) \\ U_{LQR}(k+N_c|k+N_p-2) \\ U_{DB}(k+N_p-1|\infty) \end{pmatrix}^T. \end{equation}
Assume that at least one solution such as $\mathcal{X}_k$ exists, this is, the optimization problem is feasible at $k$. Furthermore, let the control input at $k$ be given by $u(k|k)$, this is, the solution proposed by the MPC policy. Let the candidate control input at $k+1$ given by
\begin{equation} U_{k+1} = \begin{pmatrix} U_{MPC}(k|k+N_c-1) \\ U_{LQR}(k+N_c|k+N_p-1) \\ U_{DB}(k+N_p|\infty) \end{pmatrix}^T. \end{equation}
The construction of this candidate feasible solution is displayed in Figure \ref{fig:Feas}.
The following result can be therefore considered.
\begin{lemma}
Let a solution $U_k$ of the QP problem in \eqref{QP_CostFun}--\eqref{QP_DBEqConstr} that is feasible, for an initial state $x(k|k)$ and a number of iterations $N$ within the LQR that satisfy $\|x(k+N|k) - x_e(k+N|k)\|_{Q_{LQR}+K_F^T R_{LQR} K_F} \ll 1$. In the next iteration, there is at least one solution $U_{k+1}$ that is feasible, hence leading to recursive feasibility.
\end{lemma}

A Lyapunov stability analysis for $\eqref{QP_CostFun}$ (see, for instance, \cite{LIMON20082382}) shows that, if feasible, the controller is asymptotically stable. Furthermore, note that the whole optimization problem \eqref{QP_CostFun}--\eqref{QP_DBEqConstr} is reduced to a QP problem, this is, the optimization of a quadratic cost function with linear constraints. For a practical implementation, several interesting properties can be highlighted. Firstly, as \eqref{QP_DBEqConstr} guarantees the feasibility of an equilibrium trajectory for all times, constraint \eqref{QP_EqConstr} is redundant and can be ignored. Secondly, due to the linear propagation equation \eqref{QP_PropEq} and the fact that $A(k)$ and $B(k)$ are known explicitly, the state can be propagated recursively for the desired horizon, leading to one single linear transformation. In this regard, since the LQR and dead-beat controllers are explicit linear functions of the state, both propagation and control laws can be applied to reach one single constraint on state and equilibrium parameter. Consequently, a high value of the LQR horizon $N$ can be used with a very low computational cost, while $N_c$ can be as low as $3$, as considered in this work, without compromising the performance of the controller. For high horizons, specified tools can be used to reduce the number of linear constraints of the QP problem, which certainly might be linearly dependent or redundant. This allows a deep optimization of the problem prior to its resolution, raising the interest of the proposed control system for on-board systems.

\section{Case study and simulation results}\label{Sec6}
Simulation conditions are set as a near rendezvous with the Envisat spacecraft, similar to the ones considered in \cite{TubeBased}, from which the inertia tensor of the debris is considered. This is a relevant scenario within ADR in which the proposed control policy is capable of steering the chaser towards a docking point, safely avoiding collisions with the target's solar panels and antennas. The simulation initial conditions are included in Table \ref{Tab1}.
\begin{table}[!htb]
\centering
\begin{tabular}{cc}
\hline
$\bm{r}^B$ & $(1.5,2.5,1.5)^T~\mathrm{m}$ \\ \hline
$\bm{v}^L$ & $(0,0,0)^T~\mathrm{m/s}$ \\ \hline
$\prescript{B}{L}{q}$ & $(1,0,0,0)^T$ \\ \hline
$\prescript{B}{L}{\bm{\omega}}^B$ & $(0,3.53,3.53)^T~\mathrm{deg/s}$ \\ \hline
\end{tabular}
\caption{Simulation initial conditions.}
\label{Tab1}
\end{table}
As for the state and control constraints, the corresponding parameters are  $c_x=c_z=1$, $x_{min}= z_{min}=0.1~\mathrm{m}$, $u_{max}=u_{min} =0.075~\mathrm{m/s}$.
No constraints are imposed on velocity for simplicity. Note that the constraints and initial conditions are highly restrictive, allowing to evaluate the capabilities of the proposed algorithm in a meaningful and challenging scenario. For the controller design, a sampling period of $T=1\mathrm{~s}$ is used, as it is a standard value for final rendezvous phases. The MPC phase consists of only $N_c = 3$ control inputs, while the final LQR is designed considering a larger horizon of $N=30$. Both the final LQR and the MPC are tuned heuristically. A total of $35$ iterations are considered. The feasibility conditions are imposed as described in Section \ref{Sec6}, considering a simplified dead-beat constraint which leads to $y_{max} = 2.008\mathrm{~m}$ Note that a smoother, more expensive representation of the quadratic constraint for $\theta$ would increase this factor by at most a $75\%$, and might be considered if a higher feasibility region is required. Nonetheless, the obtained optimization problem is computationally cheap and well posed, being adequate for on-line applications.
\begin{figure}[htb!]
\centering
\includegraphics[width=0.45\textwidth]{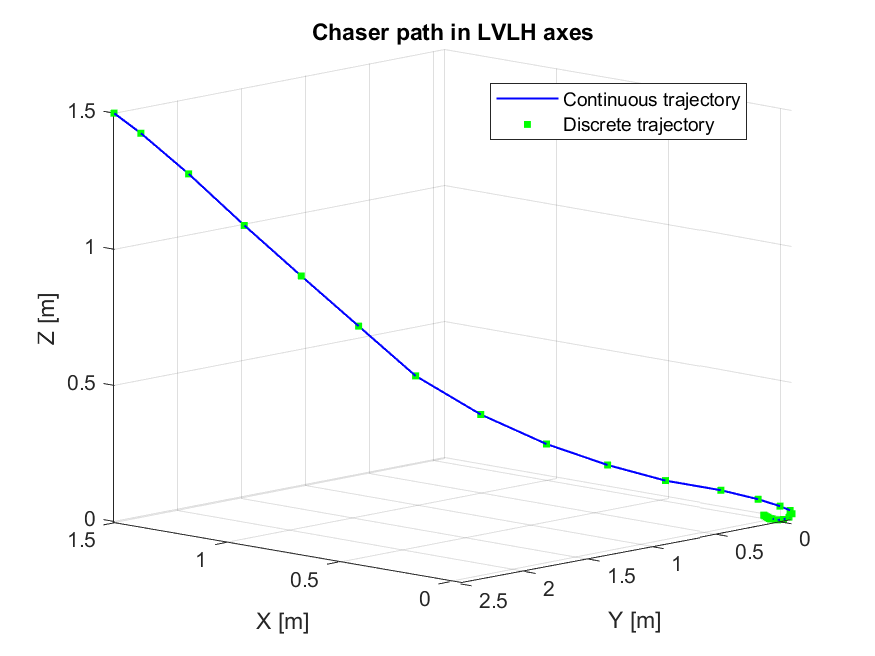}
\caption{Chaser controlled rendezvous path in LVLH axes.}
\label{fig:Fig4}
\end{figure}
\begin{figure}[htb!]
\centering
\includegraphics[width=0.45\textwidth]{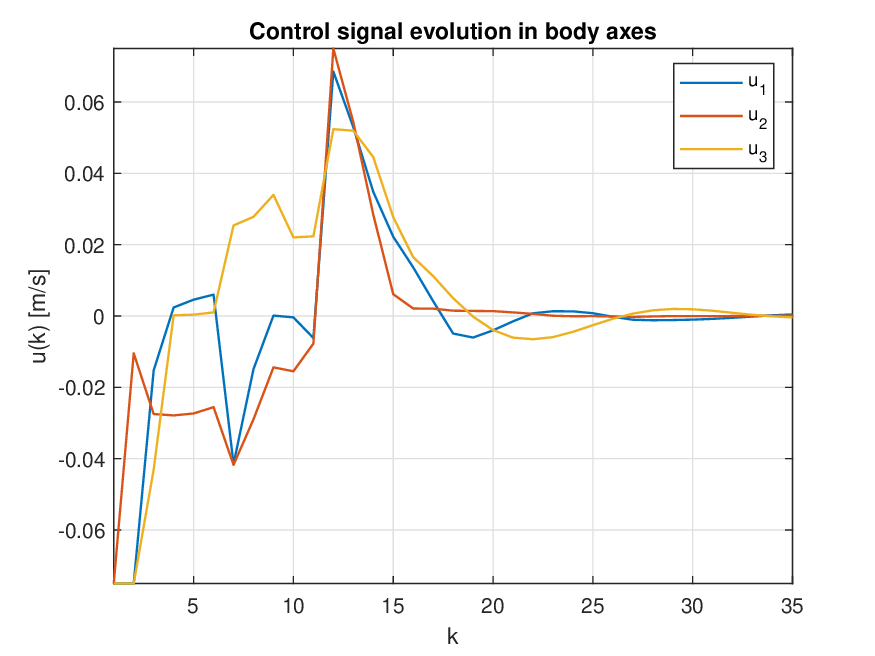}
\caption{Control signal time evolution for the controlled approach.}
\label{fig:Fig5}
\end{figure}
\begin{figure}[htb!]
\centering
\includegraphics[width=0.45\textwidth]{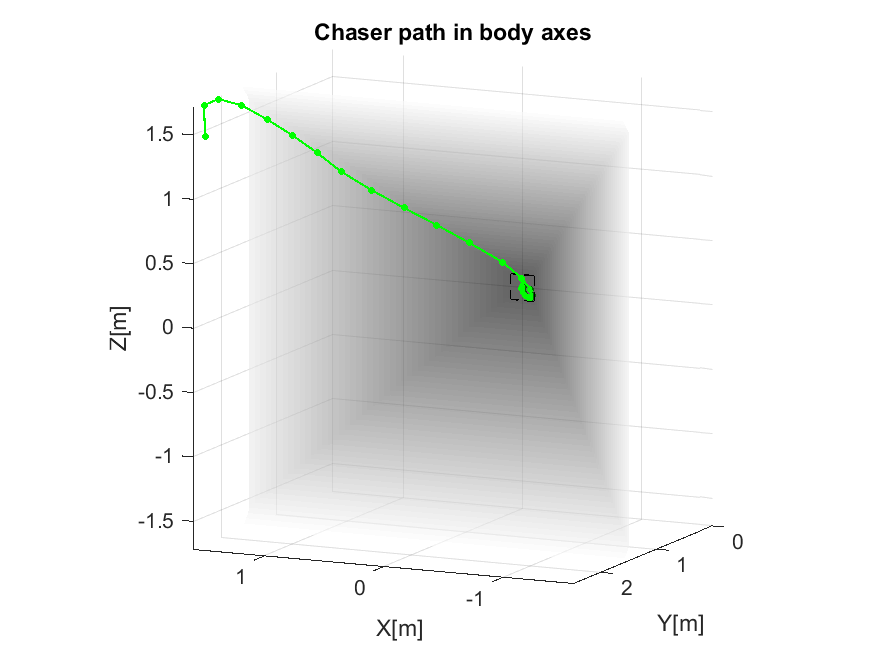}
\caption{Chaser controlled rendezvous path in body axes, including the boundaries for the LOS constraints.}
\label{fig:Fig6}
\end{figure}
The results of simulation are included in Figures \ref{fig:Fig4}--\ref{fig:Fig6}. Since the initial state of $C$ was contained within the feasible polytope of the controller, and as explicitly proved in this work, the rendezvous is guaranteed to succeed. Indeed, the control system steers the chaser to the target without any constraint violation. The required control authority is reduced, as the controller benefits from the combined HCW and target rotational dynamics, while providing strong feasibility guarantees by means of the final constraints. The domain of attraction of the proposed controller is notably enhanced by the MPCT implementation. Indeed, this same problem is tackled by using a standard MPC strategy. As to ensure a fair comparison, the same structure of \eqref{QP_CostFun}--\eqref{QP_DBEqConstr} is maintained, but imposing $\bar{\theta} = 0$ and hence removing this variable from the optimization. The problem turns infeasible until the control horizon is increased to $N_c = 35$. Comparatively, the proposed MPCT strategy makes use of an equivalent $N_c+1 = 4$ horizon, as $\bar{\theta}$ has the same dimension as $u$. Consequently, the policy developed in this work reduces drastically the number of decision variables while enhancing feasibility.

\section{Conclusions and future lines of work}\label{Sec7}
As proved throughout this paper and shown in simulation, the proposed controller successfully steers the chaser to the rotating target without any constraints violation, while allowing highly restrictive constraints in control authority by making use of the orbit and target's rotation dynamics. Moreover, the controller is asymptotically stable, locally optimal and provides recursive feasibility by design. The algorithm can be reduced to a QP optimization for each iteration, with a dimension that is the one of a conventional MPC with one more iteration in its control horizon, due to the introduction of an additional equilibrium parameter. This problem can be solved efficiently and is well suited for on-line applications. As shown in simulation, the controller far outperforms a conventional MPC implementation.

Although the controller provides highly interesting guarantees, no robust behavior was considered. Nonetheless, the adopted framework allows implementing further capabilities such as a tube-based MPC, allowing to take into consideration the uncertainties in the estimation of the target's rotational state and inertia tensor. Furthermore, a complete filtering and control algorithm which handles uncertainty granting robustness, feasibility and stability is a promising line of research. As described throughout this paper, the implementation of a general controller for elliptic orbits is natural in the chosen formulation, as it is already time-dependent and therefore allows using the discrete linearized dynamics in \cite{Yamanaka}. Finally, it is of practical interest to extend the control system to the 6DOF dynamics of the chaser, while retaining the current guarantees. The coupling between the longitudinal state of the chaser and its attitude was assessed in this work, so the addition of an attitude controller tracking the target and results on the whole 6DOF control properties are of great interest.

\section*{Funding Sources}
The authors gratefully acknowledge support by grant TED2021-132099B-C33 funded by MICIU/AEI/10.13039/501100011033 and by “European Union NextGenerationEU/PRTR.”

\bibliography{bibliography}

\appendix

\section{Supplementary Proofs}\label{Ap1}

This Appendix includes some proofs that were left out of the main text.

\begin{proof}[Proof of Lemma 1]

The fact that the dimension of $M(k)$ may change with time is undesirable, as it restricts the broader applicability of results in MPC for Tracking. This issue can be tackled, nonetheless, if the target's angular velocity is small as compared to the chaser's considered sampling rate, which is certainly a requirement for a feasible MPC implementation. For an equilibrium points analysis, it is convenient to write the HCW equations in LVLH axes. Note that the reasoning followed is also valid for other models such as the one in \cite{Yamanaka}. Therefore, starting from \eqref{HCW}, it is desirable to find a set of $\bm{x}^L(k)$, $\bm{v}^L(k)$, $\bm{u}^L(k)$ that verifies the equilibrium condition in body axes, namely
\begin{equation} A_L\begin{pmatrix} \bm{x}^L(k) \\ \bm{v}^L(k) \end{pmatrix} + B_L \bm{u}^L(k) = \begin{pmatrix} \left(\prescript{B}{L}{C}(k+1)\right)^T \prescript{B}{L}{C}(k) \bm{x}^L(k) \\ \bm{v}^L(k) \end{pmatrix}, \label{EqLVLH} \end{equation}
where a combination of rotations was used as a successive attitude matrices application. Let $\prescript{B,k}{B,k+1}{C} = \left(\prescript{B}{L}{C}(k+1)\right)^T \prescript{B}{L}{C}(k)$. For a small rotation, this matrix can be written in terms of a rotation angle $\delta\bm{\theta}(k)$, with $\|\delta\bm{\theta}(k)\|_2 \ll 1$, such that
\begin{equation} \prescript{B,k}{B,k+1}{C} \approx I_3 - \left(\delta \bm{\theta}(k)\right)^\times. \end{equation}
Substituting in \eqref{EqLVLH},
\begin{equation} A_L\begin{pmatrix} \bm{x}^L(k) \\ \bm{v}^L(k) \end{pmatrix} + B_L \bm{u}^L(k) = \begin{pmatrix} \bm{x}^L(k) \\ \bm{v}^L(k) \end{pmatrix} - \begin{pmatrix} \left(\delta \bm{\theta}(k)\right)^\times \bm{x}^L(k) \\ 0_{3\times1} \end{pmatrix}.  \end{equation}
Introducing a parameter $\theta$ and the corresponding map $M$ for equilibrium points,
\begin{equation} A_L M_x + B_L M_u = M_x - \begin{bmatrix} \left(\delta \bm{\theta}(k)\right)^\times & 0_3 \\ 0_3 & 0_3 \end{bmatrix} M_x. \label{RotEq01} \end{equation}
To simplify the problem, it is convenient to consider the separation $M  = M_0 + \Delta M$,
where
\begin{equation} \begin{bmatrix} A_L - I_6 & B_L \end{bmatrix} M_0 = 0 \leftrightarrow M_0 = \ker{\begin{bmatrix} A_L - I_6 & B_L \end{bmatrix}}. \label{SteadyKernel} \end{equation}
Note that the dimension of $M_0$ is $3$, as it is computed for the conventional HCW equations. Therefore, for consistency, this operation should be performed column by column, this is, $M_0 + \Delta M$ symbolizes the sum of any column of $M_0$ with any column of $\Delta M$, taking into account that the number of columns for these two matrices might differ. Substituting in \eqref{RotEq01},
\begin{equation} A_L \Delta M_x + B_L \Delta M_u = \Delta M_x - \begin{bmatrix} \left(\delta \bm{\theta}(k)\right)^\times & 0_3 \\ 0_3 & 0_3 \end{bmatrix} (M_{0,x} + \Delta M_x), \end{equation}
this is,
\begin{equation} \begin{aligned} \begin{bmatrix} A_L - I_6 & B_L \end{bmatrix} \Delta M = & -\begin{bmatrix} \left(\delta \bm{\theta}(k)\right)^\times & 0_3 \\ 0_3 & 0_3 \end{bmatrix} M_{0,x} \\ &-  \begin{bmatrix} \left(\delta \bm{\theta}(k)\right)^\times & 0_3 \\ 0_3 & 0_3 \end{bmatrix}  \Delta M_x. \label{RotEq02} \end{aligned} \end{equation}
Since the right-hand side of the equality is at least of order $O \left(\delta \bm{\theta}(k)\right)$ independently of $\Delta M$, as $M_0$ is of order $O \left(1\right)$, $\Delta M$ must also be $O \left(\delta \bm{\theta}(k)\right)$. Consequently, the last term in \eqref{RotEq02} is of order $O\left(\delta \bm{\theta}(k)^2 \right)$ and can be neglected. Then,
\begin{equation} \begin{bmatrix} A_L - I_6 & B_L \end{bmatrix} \Delta M = -\begin{bmatrix} \left(\delta \bm{\theta}(k)\right)^\times & 0_3 \\ 0_3 & 0_3 \end{bmatrix} M_{0,x}. \label{RotEq03} \end{equation}
$\Delta M$ can not belong to the kernel of $\begin{bmatrix} A_L - I_6 & B_L \end{bmatrix}$ because $M_{0,x}$ is nonzero. Therefore, $\Delta M$ belongs to the corresponding column space, which is of dimension $6$ as derived from the rank-nullity theorem. Equation \eqref{RotEq03} provides 6 independent equalities for $6$ degrees of freedom in $\Delta M$ and thus this perturbation matrix is completely determined. This result is of high interest to analyze the behavior of the equilibrium points in the rotating frame. Each column of $M$ is the sum of an element of the kernel and one of the column space of $\begin{bmatrix} A_L - I_6 & B_L \end{bmatrix}$.  Therefore, the columns of $M$ are linearly independent and $n_\theta = 3$.
Consequently, for a small rotation, there exists a space of equilibrium points of dimension 3 within $\mathbb{R}^{n_x + n_\theta}$. Furthermore, there exists a bijection between $\theta$ and $\bm{r}_e(k)$ for all iterations $k$. Indeed, if this bijection did not exist, the columns of $M(k)$ could be linearly combined so that at least two of the three position parts are equal, this is, in LVLH axes,
\begin{equation} M(k) = \begin{bmatrix} \bm{r}^L_{1} & \bm{r}^L_{1} & \bm{r}^L_2 \\
\prescript{B}{L}{\bm{\omega}}^L(k) \times \bm{r}_1 & \prescript{B}{L}{\bm{\omega}}^L(k) \times \bm{r}^L_1 & \prescript{B}{L}{\bm{\omega}}^L(k) \times \bm{r}^L_2 \\ \vdots & \vdots & \vdots \end{bmatrix}, \end{equation}
where the dependence of velocity with position was made explicit. However, for a given position and velocity to remain constant, there is only one possible control input, as given by \eqref{HCW}. Thus, the first two columns of $M(k)$ would be equal, which is not possible because $n_\theta(k)$ was proved to be equal to $3$ for all time steps.
\end{proof}
\hfill

\begin{proof}[Proof of Lemma 2]

Let the unit n-ball $\Omega = B^n(1)$. For a set of samples $\hat{x}_1,..., \hat{x}_l \in \Omega$, the linear constraint \eqref{BallsIneq} can be rewritten by linearity,
\begin{equation} \sum_{j=1}^{l}{A^j \theta r_j \hat{x}_j} \le b, \forall \hat{x}_j \in \Omega. \end{equation}
Consequently, the following set of independent scalar constraints must be satisfied,
\begin{equation} \theta \sum_{j=1}^{l}{\hat{A}^j_i \hat{x}_j} \le 1, \forall i \in \mathbb{N}\cap[1,m], \forall \hat{x}_j \in \Omega, \label{BallsIneq_1}\end{equation}
where, for a given matrix $A^j$, $A^j_i$ is given by its i-th row, and defining
\begin{equation} \hat{A}^j_i = A^j_i \frac{r_j}{b_i}. \end{equation}
Let the functional $J_i : \Omega\times...\times\Omega \rightarrow \mathbb{R}$,
\begin{equation} J_i(\hat{x}_1,...,\hat{x}_l) = \sum_{j=1}^{l}{\hat{A}^j_i \hat{x}_j}, \end{equation}
for with the maximum value is given by
\begin{equation} J_i^* = \max_{\hat{x}_j\in\Omega}{J_i(\hat{x}_1, ..., \hat{x}_l)}. \end{equation}
Moreover, let
\begin{equation} J^* = \max_{i \in \mathbb{N}\cap[1,m]}{J_i^*}. \end{equation}
The linear restrictions \eqref{BallsIneq_1} are trivially bounded by their maximum for a given $\theta$, namely,
\begin{equation} \theta J_i \le \max_{\hat{x}_j\in\Omega}{\theta J_i} = \theta J_i^* \le \max_{i \in \mathbb{N}\cap[1,m]}{J_i^*} = \theta J^* \le 1. \end{equation}
In consequence, the maximum admissible value of $\theta$ is given by
\begin{equation} \theta_{max} = \frac{1}{J^*}. \label{ThtMax} \end{equation}
Note that, since for each $j$, $\hat{x}_j$ are independent, from linearity,
\begin{equation} J_i^* = \max_{\hat{x}_j\in\Omega}{\sum_{j=1}^{l}{\hat{A}^j_i \hat{x}_j}} = \sum_{j=1}^{l}{\max_{\hat{x}_j \in \Omega}{\hat{A}^j_i \hat{x}_j}} = \sum_{j=1}^{l}{J_{ij}^*} \end{equation}
For a given $i\in \mathbb{N}\cap[1,m]$, the value $J_{ij}^*$ is the solution of a standard constrained optimization problem, whose solution is found in the boundary of the available set. Thus, the problem is characterized by the Lagrangian
\begin{equation} L_{ij} = \hat{A}^j_i \hat{x} + \lambda \left(\hat{x}^T \hat{x} - 1\right), \end{equation}
which can be solved explicitly giving
\begin{equation} J_{ij}^* = \sqrt{\sum_{k=1}^{n}{\left(\hat{A}^j_{ik}\right)^2}}. \end{equation}
Note that $\hat{A}^j_{ik}$ is the i-th row, k-th column element of matrix $\hat{A}^j$. Substituting the results into \eqref{ThtMax},
\begin{equation} \begin{split} \theta_{max} &= \frac{1}{\max_{i \in \mathbb{N}\cap[1,m]}{\left\{\sum_{j=1}^{l}{\sqrt{\sum_{k=1}^{n}{\left(\hat{A}^j_{ik}\right)^2}}}\right\}}} \\ &= \min_{i \in \mathbb{N}\cap[1,m]}{\left\{ \frac{b_i}{\sum_{j=1}^{l}{r_j\sqrt{\sum_{k=1}^{n}{\left(A^j_{ik}\right)^2}}}} \right\}}. \end{split} \end{equation}
\end{proof}

\begin{proof}[Proof of Lemma 3]

For this proof, several results can be considered. Firstly, the LQR control input will always be smaller in norm than the corresponding dead-beat, as the dead-beat control can be seen as an LQR where $R=0$. Therefore, if DB1 satisfies the control constraints, so does the LQR at $k+1$. Secondly, state constraints are only written for position. The position one iteration after the LQR is then the same for any control input. The position two iterations after the LQR, for the dead-beat control, is the one of the target equilibrium trajectory, which is feasible. Thus, the dead-beat is feasible for the state constraints if the LQR is so.

The missing feasibility guarantee can be obtained by exploiting the Lyapunov stability properties of the LQR. Indeed, the proposed LQR policies have their associated Lyapunov function. For simplicity, let $x_i = x^L(k+i|k) - x^L_s(k+i|k)$ and $u_i = u^L(k+i|k) - u^L_s(k+i|k)$, so that
\begin{equation} V_i = x_i^T P x_i, \end{equation}
for $P$ positive definite and given by the solution of the algebraic Riccati equation. The Lyapunov function of the LQR is related to the Hamilton-Jacobi-Bellman (HJB) equation, namely,
\begin{equation} V_k = x_k^T Q x_k + u_k^T R u_k + V_{k+1}. \end{equation}
Therefore,
\begin{equation} \begin{split} &~~u_{k+1}^T R u_{k+1} - u_k^T R u_k \\ & = 2 V_{k+1} - V_{k+2} - V_k + x_k^T Q x_k - x_{k+1}^T Q x_{k+1}. \end{split} \end{equation}
Furthermore, it holds that, for a controllable system, the infinite time LQR is exponentially stable, this is,
\begin{equation} 2 V_{k+1} - V_{k+2} - V_k = \left(V_{k+1} - V_{k+2}\right) - \left(V_k-V_{k+1}\right) < 0. \end{equation}
Thus,
\begin{equation} \left(\|x_{k+1}\|_Q^2 - \|x_k\|_Q^2\right) + \left(\|u_{k+1}\|_R^2 - \|u_k\|_R^2\right) < 0. \end{equation}
Introducing the control law $u_k = K_F x_k$,
\begin{equation} \|x_{k+1}\|_{Q + K_F^T R K_F}^2 < \|x_k\|_{Q + K_F^T R K_F}^2. \end{equation}
The obtained result is equivalent to
\begin{equation} \|x_{k+1}\|_{Q + K_F^T R K_F} < \|x_k\|_{Q + K_F^T R K_F}. \end{equation}
Let the corresponding bounding ellipsoid be given by
\begin{equation} \mathcal{X}_k = \{x \in \mathbb{R}^2 : \|x\|_{Q+K_F^T R K_F} \le \|x_k\|_{Q+K_F^T R K_F}\}. \end{equation}
As proved, $\|x_{k+i}\|_{Q+K_F^T R K_F} \in \mathcal{X}_k$. Therefore, if all $x \in \mathcal{X}_k$ verify the position constraints, the LQR is recursively feasible. Note that this condition can be effectively satisfied by using a sufficiently high number of iterations for the LQR, particularly, so that $\|x_k\|_{Q+K_F^T R K_F} \ll 1$. As this condition can be proved during the design process, the hypothesis can be indeed accepted.

The first dead-beat control is computed from a linear law,
\begin{equation} u(k+i|k) = K_{DB,x}(k+i)x(k+i|k) + K_{DB,\theta}(k+i)\bar{\theta}(k), \end{equation}
in the iterations $k+i$ that correspond. Writing the constraint on control,
\begin{equation} A_u\left(K_{DB,x}(k+i)x(k+i|k) + K_{DB,\theta}(k+i)\bar{\theta}(k)\right) < b_u. \end{equation}
The constraint on $u(k+i|k)$ is verified by explicitly requiring it in the optimization problem. The following upper bound can be established,
\begin{equation} A_u K_{DB,x}(k+i) x(k+i|k) < A_u \|K_{DB,x}(k+i)\| x(k+i|k). \end{equation}
Due to the structure of $K_{DB,x}(k+i)$, which is purely given by rotations, its norm is a constant $K_{DB,x}$ for all $k+i$. Indeed,
\begin{equation} K_{DB,x}(k+i) = \prescript{B}{L}{C}(k+i) \hat{K}_{DB} \begin{bmatrix} \prescript{B}{L}{C}(k+i)^T & 0_3 \\ 0_3 & I_3 \end{bmatrix}. \end{equation}
Furthermore, the ellipsoid $\mathcal{X}_{k+i}$ was proved to be an upper threshold for its norm in subsequent iterations. Hence, for all future iterations,
\begin{equation} \begin{split}
&A_u u_F(k+i+j|k) \le \\ \max_{\|x-x_e\| \in \mathcal{X}_{k+i} }&{A_u \left(K_{DB,x} x + A_{DB,\theta} \bar{\theta}(k)\right) }.
\end{split} \end{equation}
Provided that the tightness condition on $\mathcal{X}_k$, already written for the LQR feasibility, is satisfied, the term on $x$ can be neglected against the term on $\bar{\theta}$, resulting in a condition only for the equilibrium trajectory. As this condition is already satisfied for DB2, DB1 would then be feasible.
\end{proof}

\end{document}